\documentclass[10pt,twocolumn]{article}

\usepackage[T1]{fontenc}
\usepackage[utf8]{inputenc}
\usepackage{lmodern}
\usepackage[
  letterpaper,
  top=0.72in,
  bottom=0.78in,
  left=0.72in,
  right=0.72in,
  columnsep=0.24in
]{geometry}
\usepackage{microtype}
\usepackage{amsmath,amssymb}
\usepackage{graphicx}
\usepackage{booktabs}
\usepackage{array}
\usepackage{caption}
\usepackage{hyperref}
\usepackage{xcolor}

\graphicspath{{figures/}}
\hypersetup{
  colorlinks=true,
  linkcolor=blue!55!black,
  citecolor=blue!55!black,
  urlcolor=blue!55!black,
  pdftitle={When Firms Learn to Game the Rules},
  pdfauthor={Xufeng He}
}

\begin{document}

\twocolumn[
\begin{@twocolumnfalse}
\begin{center}
{\LARGE\bfseries When Firms Learn to Game the Rules\par}
\vspace{0.35em}
{\large An Agent-Based Reinforcement Learning Simulation of Boundary Search and Adaptive Enforcement\par}
\vspace{0.25em}
{\large Xufeng He\par}
\vspace{0.14em}
{\normalsize PERMITFOLIO Regulation Research Institute\par}
\vspace{0.14em}
{\normalsize \texttt{xufeng.he@permitfolio.org}\par}
\end{center}
\vspace{0.9em}
\noindent\textbf{Abstract.}
Rules-as-Code promises more testable legal obligations, but it also changes what regulated firms can learn. Existing work mostly emphasizes implementation gains; the strategic gap is whether machine-readable rules make boundary search cheaper. I study that gap with a synthetic agent-based reinforcement-learning simulation that separates actual conduct near a legal threshold from proximity in the computable enforcement signal. Across 150 seed-level scenario runs, 378 common-random-number computability-sweep runs, 288 Latin-hypercube global-design runs, and a 2,880,000-row firm-period panel, computable static rules raise conduct boundary mass relative to ambiguous static rules (0.411 versus 0.367) and raise signal boundary mass more sharply (0.403 versus 0.281). Ordinary adaptive updates lower consumer harm (0.202 to 0.194) but do not reliably reduce boundary search. A budget-neutral anti-gaming design reduces conduct boundary mass by 0.032 and consumer harm by 0.025 relative to computable static rules. These are mechanism-oriented synthetic results, not estimates of real firm behavior in a jurisdiction or industry. The contribution is an estimand distinction, an inspectable ABM/RL mechanism, and a reproducible artifact showing that transparent behavioral assumptions are sufficient to generate gaming-like boundary dynamics without implying that computable regulation is inherently undesirable.

\vspace{0.55em}
\vspace{1.1em}
\end{@twocolumnfalse}
]

\section{Introduction}

The promise of computable regulation is institutional rather than merely technical. If obligations are represented in machine-readable form, regulators can test compliance earlier, firms can reason over a more stable version of the rule, and agencies can reduce some of the inconsistency that comes from informal implementation. Mohun and Roberts' OECD working paper on Rules-as-Code frames this as rulemaking for both human and machine audiences \cite{oecd2020cracking}. The OECD recommendation on agile regulatory governance adds a second premise: regulation should be able to adapt as markets change, rather than rely only on slow statutory or administrative revision \cite{oecd2021agile}.

That account leaves a strategic question underdeveloped. Once firms can read the boundary more clearly, what do they learn? The relevant behavior is not only open violation. In many compliance settings the action occurs just inside the legal line: firms remain formally compliant, choose cheaper compliance forms, or discover loopholes that satisfy the coded condition while preserving privately profitable risk. A rule that is formal, testable, and predictable may reduce mistakes. It may also reduce the cost of boundary search.

The research question of this paper is therefore: when regulatory obligations become more computable, do firms cluster closer to the legal boundary, and can adaptive anti-gaming design reduce the resulting harm without simply increasing audit volume? The answer requires separating two things that are often blended together. A more computable rule can make the enforcement signal cleaner even if underlying conduct does not change. Conversely, firms may actually move closer to the threshold while the measured signal makes that movement easier to see.

This mechanism is related to Goodhart's law, but the legal setting is not a generic metric-gaming problem \cite{goodhart1975problems}. Legal rules create administrable boundaries, not only performance scores. Firms observe one another and can copy profitable edge strategies. Regulators can move thresholds, change audit intensity, randomize enforcement margins, or add outcome-based backstops. The object of analysis is a feedback system: rule formalization, firm learning, detection, regulatory response, and renewed search.

I use a synthetic simulation because the target mechanism is difficult to observe cleanly in one jurisdiction or sector. The model does not estimate a real market. It makes a set of behavioral assumptions explicit enough to inspect: firms learn, competitors imitate, consumers respond to perceived risk, and regulators revise rules under noisy information. This generative use of simulation follows the logic of agent-based modeling: the point is not to fit one dataset, but to test whether a proposed mechanism can produce the pattern of interest under transparent assumptions \cite{epstein2006generative,tesfatsion2006handbook}.

The paper makes three contributions. The theoretical contribution is an estimand distinction between conduct boundary mass and signal boundary mass. The methodological contribution is a reproducible simulation environment that combines firm reinforcement learning, competitor imitation, common-random-number sweeps, paired sign-flip inference, de-overlapped event studies, and firm-period transition analysis. The computational contribution is synthetic but empirically disciplined through seed-level inference, sensitivity analysis, and firm-period microdata: under the specified behavioral assumptions, computability increases both signal clustering and conduct clustering; ordinary adaptation lowers harm without reliably reducing boundary mass; budget-neutral anti-gaming lowers harm and conduct boundary mass but does not eliminate edge strategies.

The main result is narrower than the strongest gaming story, but more defensible. Computable static rules raise conduct boundary mass from 0.367 to 0.411, while signal boundary mass rises from 0.281 to 0.403. Ordinary adaptive regulation lowers consumer harm from 0.202 to 0.194, but its paired conduct-boundary effect is small (-0.002) and statistically unstable. Anti-gaming adaptive design reduces harm by 0.025 and boundary mass by 0.032. It does not remove edge strategies: their late-period share moves from 0.868 to 0.802, while the intervention-trigger rate moves from 0.041 to 0.098. A naive reinforcement-learning regulator reduces harm, but it produces high rule churn (3.111). Learning capacity, by itself, is not institutional design.

\noindent\textbf{Scope and contribution.}
The paper makes a mechanism claim rather than a jurisdiction-specific estimate, a legal-advice document, or a prescriptive regulatory blueprint. Under explicit behavioral assumptions, computable rules can change both conduct and enforcement signals; simple adaptive updating is not the same as anti-gaming design; and a reusable artifact can make those distinctions inspectable.

\section{Research Debate and Hypotheses}

The literature on computable regulation pulls in two directions. The implementation view treats machine-readable rules as institutional infrastructure: legal requirements become easier to test, automate, update, and communicate to both officials and regulated parties \cite{oecd2020cracking,oecd2021agile}. That view is compelling, especially in domains where legal ambiguity produces inconsistent advice or expensive manual screening. But it tends to assume that better rule observability is used mainly for compliance.

A second line of work gives a less comfortable prediction. When indicators, scores, or formal tests become targets, actors may optimize against the measure rather than the underlying public purpose \cite{goodhart1975problems,campbell1979assessing}. Law adds a complication that generic metric-gaming accounts often miss: legal rules do not only measure performance; they create administrable lines. Code can therefore operate as regulation in Lessig's sense, while rule-based legality still invites strategic line-walking in Schauer's sense \cite{lessig1999code,schauer1991playing}. The unresolved question is whether computability mainly clarifies legal obligations or also makes the legal edge cheaper to find.

I treat computable regulation as an information structure. A less computable rule leaves firms uncertain about where the boundary lies. A highly computable rule makes the enforcement condition more reproducible. That can reduce accidental noncompliance, but it also lowers the cost of strategic search. The key move is to separate observed proximity to the computable signal from actual proximity in conduct. Without that separation, a cleaner measurement system can be mistaken for a behavioral shift, or a behavioral shift can be dismissed as mere measurement precision.

Let $\theta_t$ denote the operative legal threshold at time $t$, and let $r_{it}$ denote firm $i$'s conduct risk. A firm is formally compliant when $r_{it} \leq \theta_t$. Boundary compliance is the event $0 \leq \theta_t-r_{it}\leq \epsilon$, where $\epsilon$ is a small margin just inside the legal line. Conduct boundary mass is the population share of firms in that region:
\[
B_t=\frac{1}{N}\sum_i \mathbf{1}\{0 \leq \theta_t-r_{it}\leq \epsilon\}.
\]

Four hypotheses follow from this setup.

\noindent\textbf{H1.} Computability increases signal boundary mass by making the enforcement signal more precise; it increases conduct boundary mass only if strategic search dominates measurement clarification.

\noindent\textbf{H2.} Competitor imitation may amplify profitable edge strategies. The model treats this as an empirical question rather than a guaranteed direction: imitation may appear as boundary clustering, loophole migration, or higher harm, depending on which strategies remain profitable.

\noindent\textbf{H3.} Ordinary adaptive updating can create a regulatory-arbitrage cycle. If regulators update thresholds after observing harm or clustering, firms may continue boundary-oriented search around the new rule rather than necessarily returning to thicker substantive compliance.

\noindent\textbf{H4.} Anti-gaming design reduces the double edge mainly by reducing harm and making edge behavior less attractive at the boundary; reductions in the narrower loophole-shift share are a separate and weaker empirical claim. The model expects guardrail effects to be larger than randomization effects when harmful edge conduct is observable.

\begin{figure*}[t]
\centering
\includegraphics[width=0.94\textwidth]{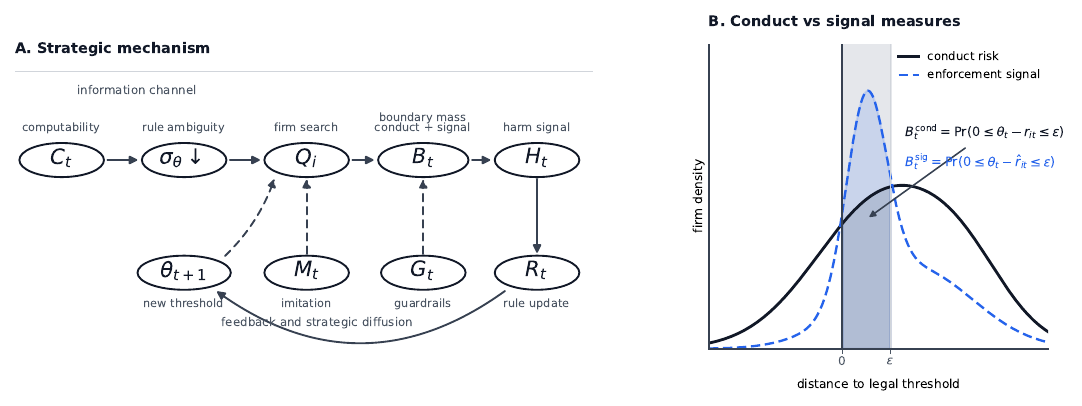}
\caption{Strategic mechanism and estimand. Panel A shows how computable rules reduce rule ambiguity, can accelerate firm search near the legal boundary, and create feedback through imitation, guardrails, harm signals, and rule updates. Panel B separates the threshold band used for conduct boundary mass from the corresponding band in the computable enforcement signal.}
\label{fig:mechanism}
\end{figure*}

\section{Model}

The model has four agent classes: firms, a regulator, consumers, and a competitor-imitation network. Firm agents choose among seven strategies, from quality overcompliance to open noncompliance. Each action carries a compliance cost, price effect, latent harm level, loophole intensity, intended compliance margin, adjustment speed, and adjustment cost. These numbers are not estimated from a real sector. They are design parameters chosen to make the mechanism inspectable: moving from overcompliance to edge behavior lowers private cost and can raise latent harm, while not every low-cost action is automatically illegal.

One modeling choice matters more than the others. Firms hold an absolute conduct-risk state with threshold-relative target adjustment. Each strategy defines a target margin around the operative threshold, but actual conduct risk adjusts gradually and does not instantaneously move with the threshold. The observed enforcement signal, by contrast, becomes more precise as rules become more computable. I adopted this separation after checking an earlier threshold-relative construction: in that version, lower signal noise could mechanically create apparent boundary clustering even without meaningful behavioral movement. The present estimand is meant to avoid that artifact.

Firms update action values with tabular Q-learning \cite{sutton2018reinforcement}. I use this deliberately simple learning rule because the paper's claim depends on inspecting the mechanism, not on maximizing predictive performance. The firm's reward is profit net of compliance cost, expected penalty, adjustment cost, and reputational damage. The state representation discretizes the latest signal-based distance to the boundary, enforcement pressure, and recent market harm. This is a coarse representation, but it keeps the policy space auditable.

Regulators choose among static rules, periodic adaptive updates, Q-learning threshold updates, and anti-gaming updates. Their loss increases with consumer harm, audit cost, observed boundary pressure, and rule churn. Consumers allocate demand using price, perceived risk, reputation, and quality. A competitor-imitation channel links firms to one another: high-profit strategies are copied with a probability that rises as rule computability increases.

Computability is modeled as a bundle because real Rules-as-Code deployments rarely change only one thing. They can clarify thresholds, make scoring machine-readable, reduce signal noise, speed firm adjustment, make enforcement pressure more legible, and help strategies diffuse. I therefore do not claim component-level causal identification in the core scenarios. The common-random-number sweep and Latin-hypercube analysis test whether the pattern survives parameter movement, while the channel ablation in Table~\ref{tab:diagnostics} is a diagnostic stress test rather than an estimate of separate real-world technology effects.

\begin{table*}[t]
\centering
\caption{Firm action parameters. These design parameters define the payoff and harm structure used in the synthetic mechanism model.}
\label{tab:actions}
\footnotesize
\setlength{\tabcolsep}{3pt}
\resizebox{\textwidth}{!}{%
\begin{tabular}{lrrrrrrrr}
\toprule
Action & Margin & Cost & Price disc. & Latent harm & Loophole & Quality & Adj. speed & Adj. cost\\
\midrule
quality\_overcompliance & 0.220 & 0.210 & -0.020 & 0.030 & 0.000 & 0.920 & 0.340 & 0.200\\
ordinary\_compliance & 0.120 & 0.145 & 0.000 & 0.060 & 0.000 & 0.720 & 0.300 & 0.160\\
lean\_compliance & 0.070 & 0.105 & 0.040 & 0.085 & 0.050 & 0.610 & 0.270 & 0.130\\
boundary\_test & 0.028 & 0.065 & 0.080 & 0.125 & 0.180 & 0.480 & 0.240 & 0.100\\
aggressive\_edge & 0.008 & 0.042 & 0.110 & 0.165 & 0.300 & 0.380 & 0.220 & 0.080\\
loophole\_shift & 0.040 & 0.052 & 0.100 & 0.205 & 0.700 & 0.350 & 0.180 & 0.060\\
open\_noncompliance & -0.030 & 0.018 & 0.130 & 0.265 & 0.300 & 0.220 & 0.120 & 0.040\\
\bottomrule
\end{tabular}}
\end{table*}

The primary outcomes are conduct boundary mass, signal boundary mass, consumer harm, edge-strategy share, loophole-shift share, formal violation rate, threshold detections, guardrail triggers, intervention triggers, rule churn, and a cycle index that combines threshold volatility, audit-rate movement, harm volatility, and update frequency. I treat the cycle index as a diagnostic, not as a welfare measure. It is meant to flag regulatory-arbitrage dynamics that can be missed when attention stays on violation counts alone.

\section{Experimental Design}

The simulation compares five institutional regimes. Ambiguous static rules have low computability ($C=0.25$) and no update. Computable static rules have high computability ($C=0.85$) and no update. Computable adaptive rules update the threshold and audit rate when observed harm or signal boundary clustering rises. The RL regulator uses tabular Q-learning over five regulator actions: hold, tighten, relax, increase audits, and decrease audits. Anti-gaming adaptive design adds randomized audit margins and an outcome guardrail based on a noisy, partly delayed harm signal. Its expected audit probability is normalized to the current audit-rate budget each period, so the main treatment reallocates audit attention rather than raising expected audit volume; the audit-capacity-only case is reported separately as an ablation.

Each core regime uses 30 matched random seeds (100--129), 240 periods per seed, and 80 firms per market. The boundary band is fixed at $\epsilon=0.045$ around an initial threshold of 0.58. Scenario summaries use the final 30\% of each run, rather than the full trajectory, because early periods are dominated by exploration and initial adjustment. A computability sweep varies rule computability from 0.10 to 0.92 for static, adaptive, and anti-gaming designs under common random numbers. Robustness checks vary imitation strength, consumer risk aversion, firm learning rate, and target harm. The ablations remove the anti-gaming components one at a time, which separates audit randomization from outcome-based backstops.

The experimental package adds four checks that became necessary during model auditing. A 96-point Latin-hypercube design, repeated across three seeds, jointly varies computability, imitation, firm learning rate, consumer risk aversion, target harm, randomized audit margin, and outcome guardrail strength; standardized sensitivity is estimated on design-level means rather than treating seed replicates as independent parameter draws. Paired seed-level sign-flip tests compare each adaptive regime to computable static rules without assuming normality. The event study aligns saved time series around actual rule changes, defined as nonzero threshold or audit-rate movement, and drops nearby repeat events within the same seed. A 2,880,000-row firm-period panel records every observed firm action, conduct risk score, signal risk score, enforcement score, distance to boundary, profit, threshold detection, guardrail trigger, intervention trigger, and regulator action, which supports the transition-matrix and endgame analyses.

There is no missing data in the usual empirical sense because the panel is generated by simulation. The relevant data-cleaning decisions are instead definitional: using the operative threshold for each period, distinguishing formal violations from intervention triggers, separating threshold detections from guardrail triggers, and reporting design-level means for the Latin-hypercube sensitivity analysis. I also keep the full firm-period panel rather than only seed-level summaries so that strategy persistence can be inspected directly.

The design is mechanism-oriented: it evaluates whether plausible behavioral assumptions are sufficient to generate boundary clustering, loophole migration, and regulatory update cycles. Its empirical role is to make that mechanism inspectable and reproducible rather than to estimate a particular jurisdiction or market.

\section{Results}

\subsection{Static computability increases both conduct and signal boundary mass}

The static comparison gives the cleanest test of H1 because the regulator does not move the threshold. Conduct boundary mass is 0.367 under ambiguous static rules and 0.411 under computable static rules. Signal boundary mass also rises, from 0.281 to 0.403. The larger signal response is expected: a more reproducible enforcement score places observations more sharply around the threshold. The more important finding is that the conduct measure also moves after the model separates actual risk from the enforcement signal. That does not prove real-world boundary gaming, but it shows that the mechanism is not only a measurement artifact under the stated assumptions. Consumer harm rises from 0.175 to 0.202, which is consistent with firms finding profitable edge strategies rather than merely becoming better informed about safe compliance.

\begin{table*}[t]
\centering
\caption{Scenario-level results. Values are rounded tail-period means across 30 seeds. Paired effects in Table~\ref{tab:tests} are computed within matched seeds, so they need not equal the rounded differences between scenario means exactly.}
\label{tab:scenario}
\setlength{\tabcolsep}{3pt}
\resizebox{\textwidth}{!}{%
\begin{tabular}{lcccccccccc}
\toprule
Scenario & Conduct $B$ & Signal $B$ & Harm & Edge & Loophole-shift & Formal viol. & Threshold detect & Guardrail & Intervention & Churn\\
\midrule
Ambiguous static rules & 0.367 & 0.281 & 0.175 & 0.830 & 0.145 & 0.223 & 0.046 & 0.000 & 0.046 & 0.000\\
Computable static rules & 0.411 & 0.403 & 0.202 & 0.868 & 0.114 & 0.271 & 0.041 & 0.000 & 0.041 & 0.000\\
Computable adaptive rules & 0.409 & 0.399 & 0.194 & 0.843 & 0.122 & 0.224 & 0.058 & 0.000 & 0.058 & 0.244\\
RL regulator & 0.382 & 0.375 & 0.183 & 0.810 & 0.133 & 0.185 & 0.088 & 0.000 & 0.088 & 3.111\\
Adaptive anti-gaming design & 0.380 & 0.373 & 0.177 & 0.802 & 0.126 & 0.211 & 0.037 & 0.092 & 0.098 & 0.200\\
\bottomrule
\end{tabular}}
\end{table*}

\begin{figure}[t]
\centering
\includegraphics[width=\columnwidth]{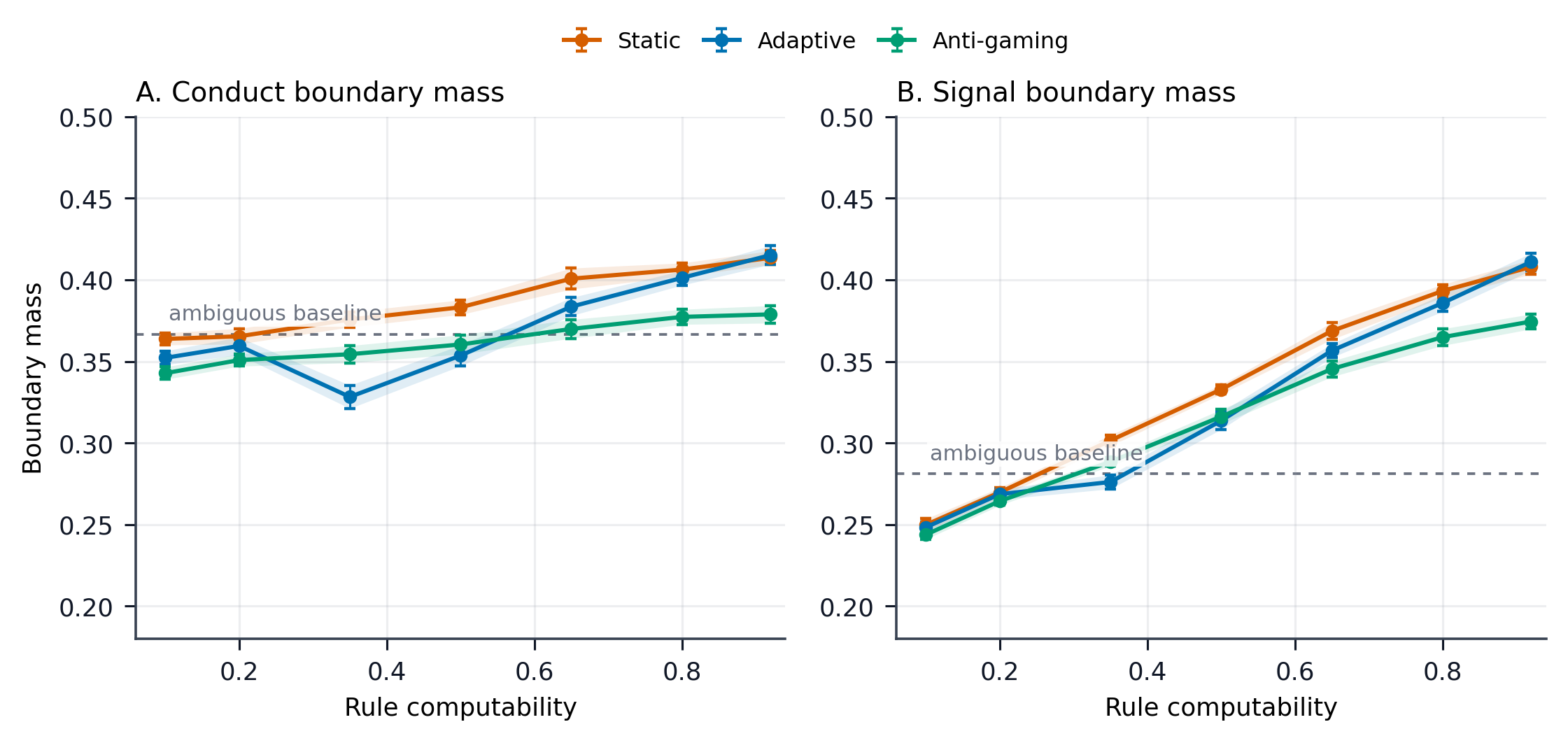}
\caption{Conduct and signal boundary mass across the computability sweep.}
\label{fig:boundary-sweep}
\end{figure}

\subsection{Ordinary adaptation lowers harm but adds churn}

Ordinary adaptation improves one outcome while adding institutional movement. Mean consumer harm falls to 0.194. Conduct boundary mass is 0.409; in the paired seed comparison with computable static rules, the effect is -0.002, with confidence intervals that cross zero in the full test table. The evidence therefore supports a limited claim. Adaptive updates reduce damage in this model, but deterministic threshold movement does not reliably reduce boundary search. A reviewer could reasonably ask whether a better adaptive rule would perform differently; the point here is that faster updating alone is not enough.

\begin{figure}[t]
\centering
\includegraphics[width=\columnwidth]{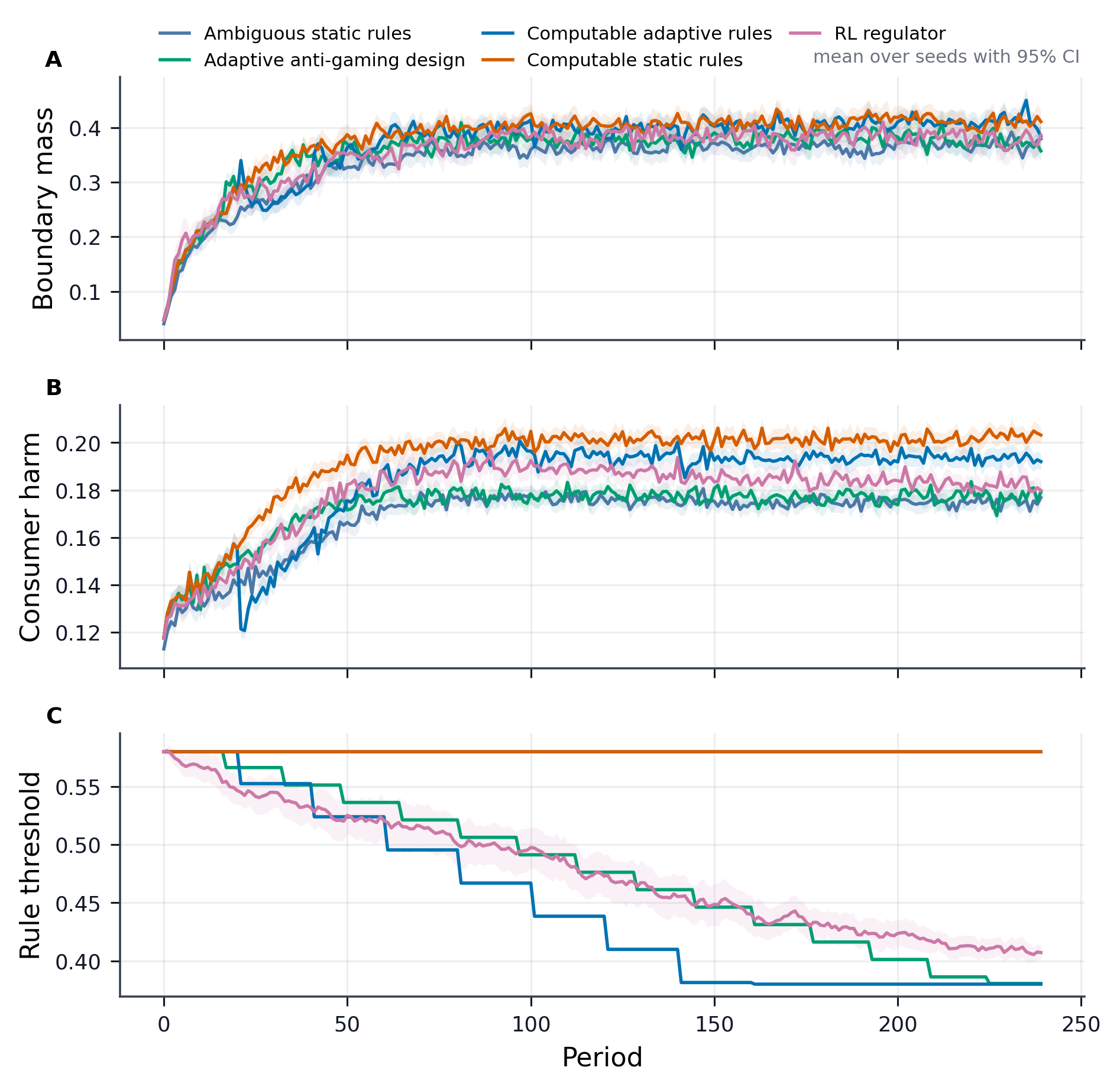}
\caption{Seed-averaged learning dynamics under alternative regulatory regimes.}
\label{fig:timeseries}
\end{figure}

\subsection{Anti-gaming design changes the equilibrium}

Anti-gaming adaptation changes the equilibrium mainly by reducing harm and boundary mass under the same expected audit budget. Conduct boundary mass falls to 0.380, and consumer harm falls to 0.177. Edge-strategy share does not disappear; it moves from 0.868 under computable static rules to 0.802. The narrower loophole-shift action moves from 0.114 to 0.126. In the paired seed tests, the reduction relative to computable static rules is 0.032 for boundary mass and 0.025 for harm, while edge-strategy share changes by -0.066. Intervention triggers rise from 0.041 to 0.098. The resulting interpretation is that anti-gaming design changes which edge strategies remain profitable and which become more likely to trigger review.

The ablation results show where the improvement comes from. Because the core anti-gaming regime is budget-neutral in expected audit probability, the audit-capacity-only row is a separate comparison rather than part of the treatment bundle. Removing the outcome guardrail raises harm more than removing the randomized margin. Randomization has a smaller and less stable effect on conduct boundary mass, while noisy outcome guardrails do more of the harm reduction. The advantage is conditional: it is strongest when the regulator's harm target remains demanding and when harmful edge conduct is observable enough for the guardrail to bind.

\begin{figure}[t]
\centering
\includegraphics[width=\columnwidth]{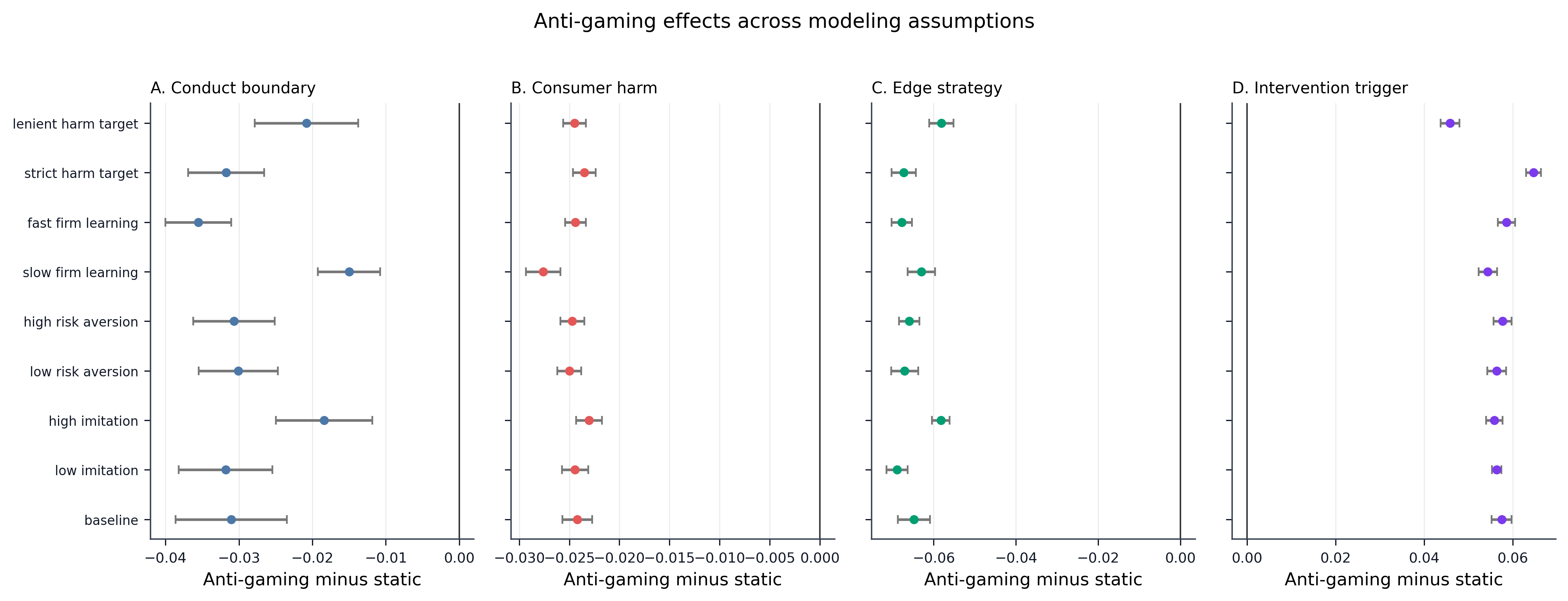}
\caption{Anti-gaming effects on boundary mass, harm, edge strategies, and intervention triggers.}
\label{fig:robustness}
\end{figure}

\begin{table}[t]
\centering
\caption{Anti-gaming ablation. Values use the 12-seed ablation sample, not the 30-seed core scenario sample in Table~\ref{tab:scenario}.}
\label{tab:ablation}
\footnotesize
\setlength{\tabcolsep}{2.4pt}
\resizebox{\columnwidth}{!}{%
\begin{tabular}{@{}lccccccc@{}}
\toprule
Design & $B$ & Harm & Edge & Loophole & Audit prob. & Intervention & Churn\\
\midrule
Adaptive anti-gaming & 0.383 & 0.177 & 0.800 & 0.128 & 0.120 & 0.099 & 0.200\\
Audit capacity only & 0.408 & 0.196 & 0.852 & 0.119 & 0.160 & 0.052 & 0.200\\
No outcome guardrail & 0.397 & 0.197 & 0.859 & 0.119 & 0.120 & 0.047 & 0.200\\
No randomized margin & 0.389 & 0.179 & 0.809 & 0.125 & 0.120 & 0.099 & 0.200\\
\bottomrule
\end{tabular}}
\end{table}

\subsection{Learning capacity is not enough}

The RL regulator is useful as a diagnostic case. It lowers harm and boundary mass relative to computable static rules, but it produces the highest rule churn (3.111). I do not read this as evidence against reinforcement learning in regulation. A short-horizon Q-learning regulator may simply be undertrained, and its reward function does not fully encode institutional stability, contestability, or anti-gaming structure. The result is therefore a warning about objective design and convergence, not a claim that regulatory learning is inherently unstable.

\begin{figure}[t]
\centering
\includegraphics[width=\columnwidth]{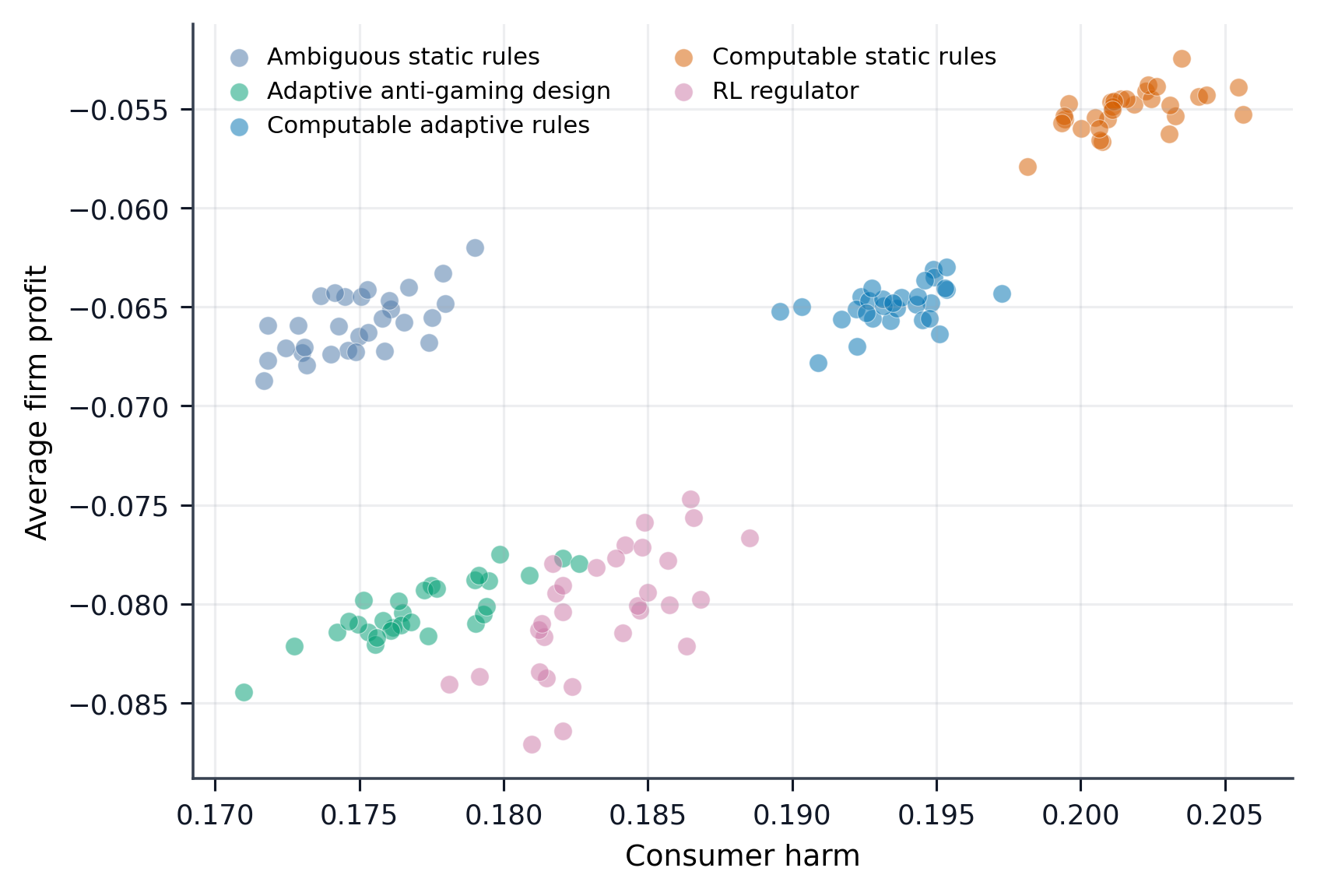}
\caption{Private profit and public harm across simulated markets.}
\label{fig:tradeoff}
\end{figure}

\begin{figure}[t]
\centering
\includegraphics[width=0.92\columnwidth]{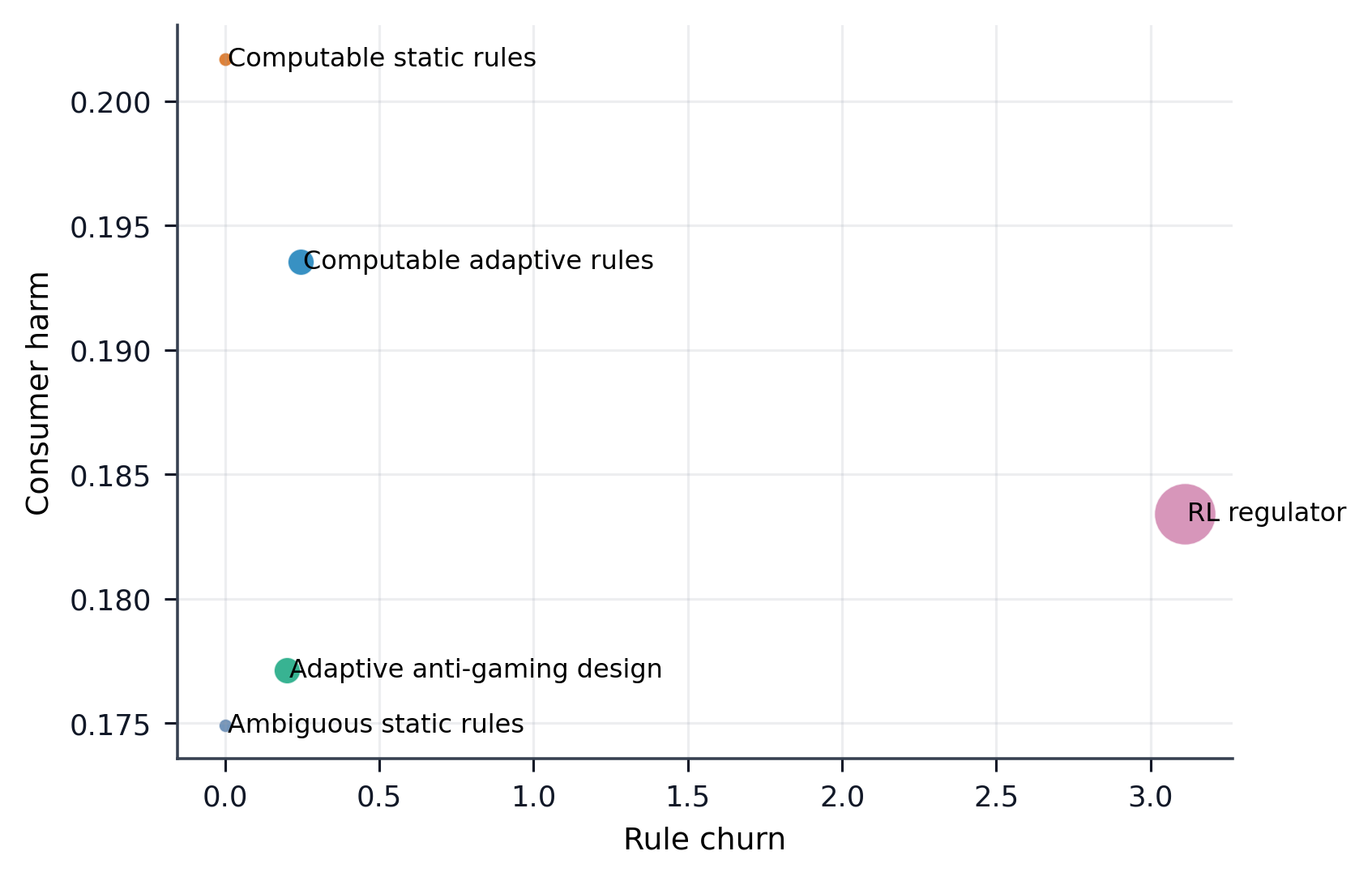}
\caption{Rule churn and consumer harm.}
\label{fig:churn}
\end{figure}

\subsection{Standardized linear sensitivity and policy frontier}

The Latin-hypercube runs make the signal-conduct distinction hard to ignore. Computability is a strong positive predictor of signal boundary mass ($\beta=0.972$, $R^2=0.978$) and also predicts conduct boundary mass in the global design ($\beta=0.780$, $R^2=0.749$). Imitation is positive but smaller for conduct boundary mass ($\beta=0.105$). Consumer harm is shaped by computability, imitation, firm learning rate, target harm, and outcome guardrails; the coefficient on outcome guardrails is -0.144, and the consumer-harm model has $R^2=0.933$. These coefficients are sensitivity diagnostics within the simulated design space and indicate which parts of the mechanism carry the largest modeled effects.

Three additional diagnostics are the main defense against overreading the simulation. Signal precision alone is not enough to reproduce the conduct result: it generates a large signal-boundary response (0.370), but conduct boundary mass remains 0.376 rather than the full-bundle value of 0.413. The boundary-epsilon check also cuts both ways. The computable-minus-ambiguous conduct gap is positive from $\epsilon=0.025$ through $\epsilon=0.080$, but the anti-gaming boundary reduction fades at the widest band because the metric then includes thicker compliance behavior, not only edge search. Penalty and reputation checks address a different objection, namely that open noncompliance might be an artifact of weak sanctions. Under a high-penalty variant, computability still raises conduct boundary mass (+0.045), while anti-gaming lowers harm (-0.031) and reduces open noncompliance from 0.485 to 0.310. These checks strengthen the mechanism claim, but they also define its boundary: the model supports a cautious claim about sufficient conditions, not a universal prediction.

\begin{table*}[t]
\centering
\caption{Mechanism and robustness diagnostics. Channel ablations use 12 seeds under computable static rules. Epsilon rows recompute boundary mass from the firm-period panel. Penalty rows use 12-seed diagnostic variants.}
\label{tab:diagnostics}
\footnotesize
\setlength{\tabcolsep}{4pt}
\begin{tabular}{llccc}
\toprule
Diagnostic & Setting & Conduct result & Signal result & Harm or open-noncompliance result\\
\midrule
Channel & Full bundle & $B=0.413$ & $S=0.404$ & Harm $=0.201$\\
Channel & Signal precision only & $B=0.376$ & $S=0.370$ & Harm $=0.179$\\
Channel & Full except imitation & $B=0.405$ & $S=0.395$ & Harm $=0.192$\\
Channel & Full except adjustment speed & $B=0.400$ & $S=0.392$ & Harm $=0.203$\\
Epsilon & $\epsilon=0.025$ & Comp.-ambig. $=+0.035$ & Comp.-ambig. $=+0.073$ & Anti-static $B=-0.028$\\
Epsilon & $\epsilon=0.045$ & Comp.-ambig. $=+0.045$ & Comp.-ambig. $=+0.122$ & Anti-static $B=-0.032$\\
Epsilon & $\epsilon=0.080$ & Comp.-ambig. $=+0.012$ & Comp.-ambig. $=+0.137$ & Anti-static $B=+0.001$\\
Penalty & Low penalty & Comp.-ambig. $=+0.026$ & -- & Open static/anti $=0.555/0.429$\\
Penalty & High penalty & Comp.-ambig. $=+0.045$ & -- & Open static/anti $=0.485/0.310$\\
\bottomrule
\end{tabular}
\end{table*}

The non-dominated frontier contains 10 seed-level policies when conduct boundary mass, edge-strategy share, consumer harm, and rule churn are minimized jointly. This is not a dominance proof for anti-gaming design. Some ambiguous static runs remain on the frontier because they have zero rule churn. Among computable regimes, however, frontier points are concentrated in anti-gaming runs. That is the tradeoff the model is meant to make visible: computability can be valuable, but its institutional design determines whether it becomes a stable enforcement aid or a target for strategic search.

\begin{table*}[t]
\centering
\caption{Standardized linear sensitivity coefficients.}
\label{tab:sensitivity}
\footnotesize
\begin{tabular}{llcc}
\toprule
Outcome & Predictor & Standardized coefficient & $R^2$\\
\midrule
Conduct boundary & Computability & 0.780 & 0.749\\
Conduct boundary & Target harm & 0.240 & 0.749\\
Conduct boundary & Outcome guardrail & -0.125 & 0.749\\
Conduct boundary & Imitation base & 0.105 & 0.749\\
Signal boundary & Computability & 0.972 & 0.978\\
Signal boundary & Target harm & 0.068 & 0.978\\
Signal boundary & Imitation base & 0.037 & 0.978\\
Signal boundary & Consumer risk aversion & 0.032 & 0.978\\
Consumer harm & Computability & 0.769 & 0.933\\
Consumer harm & Firm learning rate & -0.392 & 0.933\\
Consumer harm & Imitation base & 0.352 & 0.933\\
Consumer harm & Outcome guardrail & -0.144 & 0.933\\
Edge strategy & Computability & 0.545 & 0.884\\
Edge strategy & Imitation base & 0.411 & 0.884\\
Edge strategy & Firm learning rate & -0.379 & 0.884\\
Edge strategy & Target harm & 0.352 & 0.884\\
\bottomrule
\end{tabular}
\end{table*}

\begin{figure}[t]
\centering
\includegraphics[width=\columnwidth]{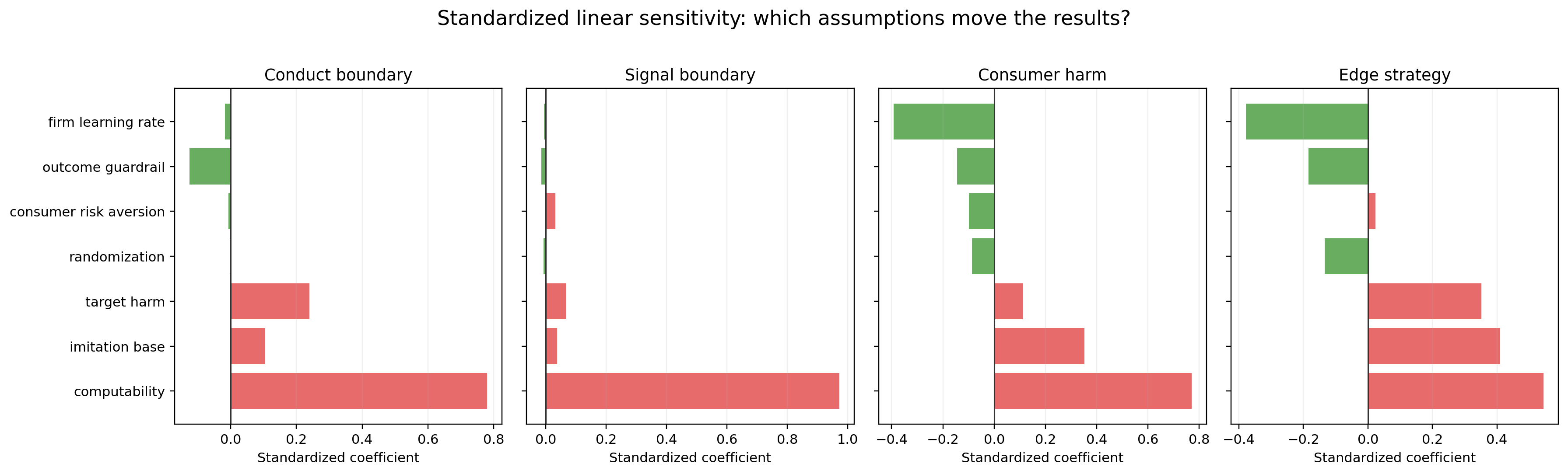}
\caption{Standardized linear sensitivity coefficients for conduct boundary mass, signal boundary mass, consumer harm, and edge-strategy share.}
\label{fig:sensitivity}
\end{figure}

\begin{figure}[t]
\centering
\includegraphics[width=0.94\columnwidth]{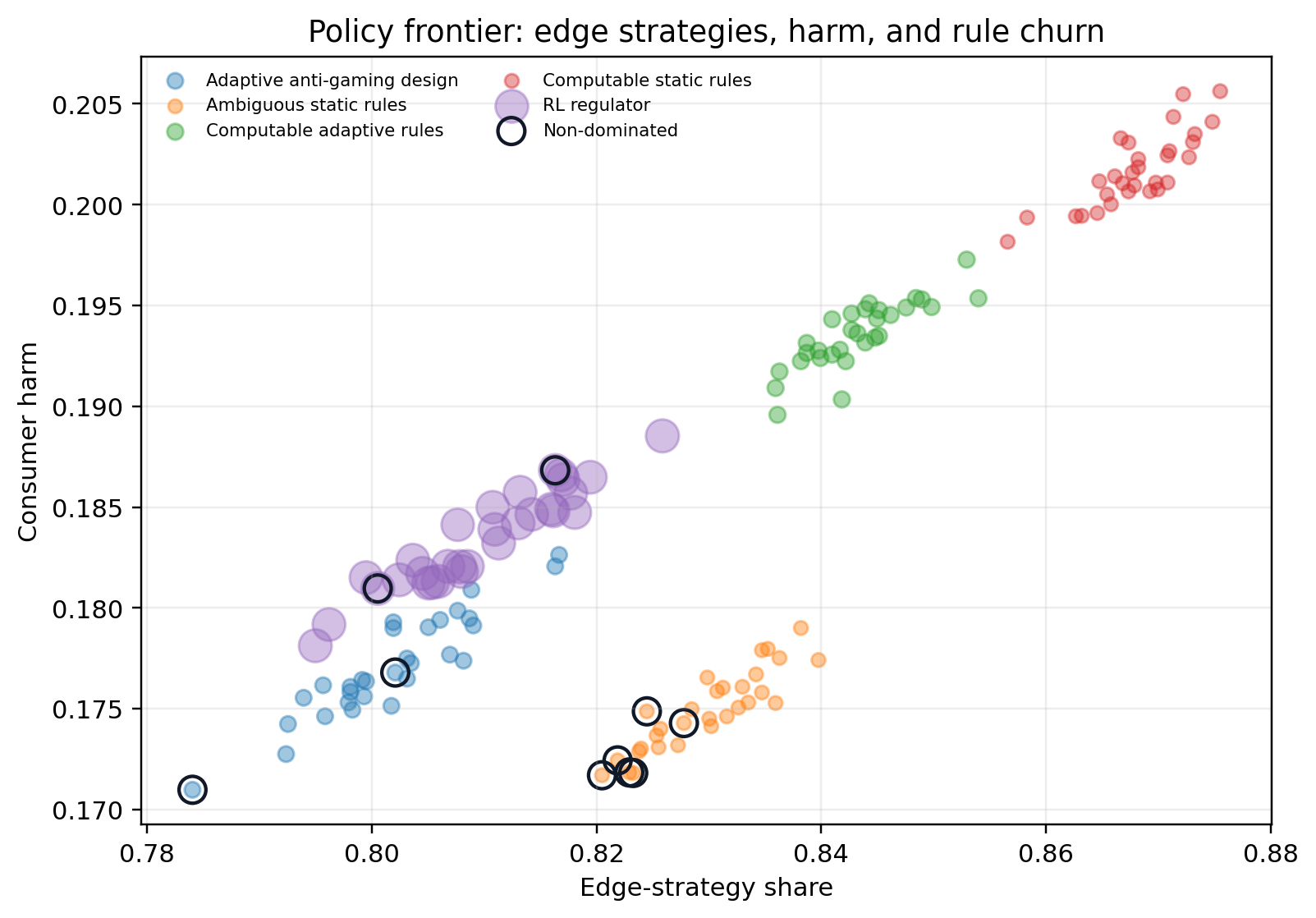}
\caption{Policy frontier for harm, edge strategies, boundary mass, and rule churn. Point size scales with rule churn; circles mark non-dominated runs.}
\label{fig:frontier}
\end{figure}

\subsection{Event-study and micro-transition evidence}

The event study aligns simulation periods around actual rule changes rather than every nominal regulator action. For anti-gaming updates, the post-update minus pre-update change in mean consumer harm is 0.002. The period-by-period effect is small, which is useful rather than disappointing: it warns against treating rule revision as an immediate reset. Ordinary adaptive and RL updates mostly move the threshold and audit rate without breaking the boundary-learning response. Firms do not become substantively compliant just because the rule changes; many relocate around the next computable boundary.

The firm-period panel gives a more granular view of the same process. Under computable static rules, late-simulation behavior concentrates in open noncompliance, loophole shifting, and boundary testing. Under anti-gaming design, late behavior shifts toward lean and ordinary compliance, although boundary testing does not disappear. The transition matrices matter here because they show persistence, not just averages: anti-gaming design changes how long harmful edge strategies remain attractive.

\begin{figure}[t]
\centering
\includegraphics[width=\columnwidth]{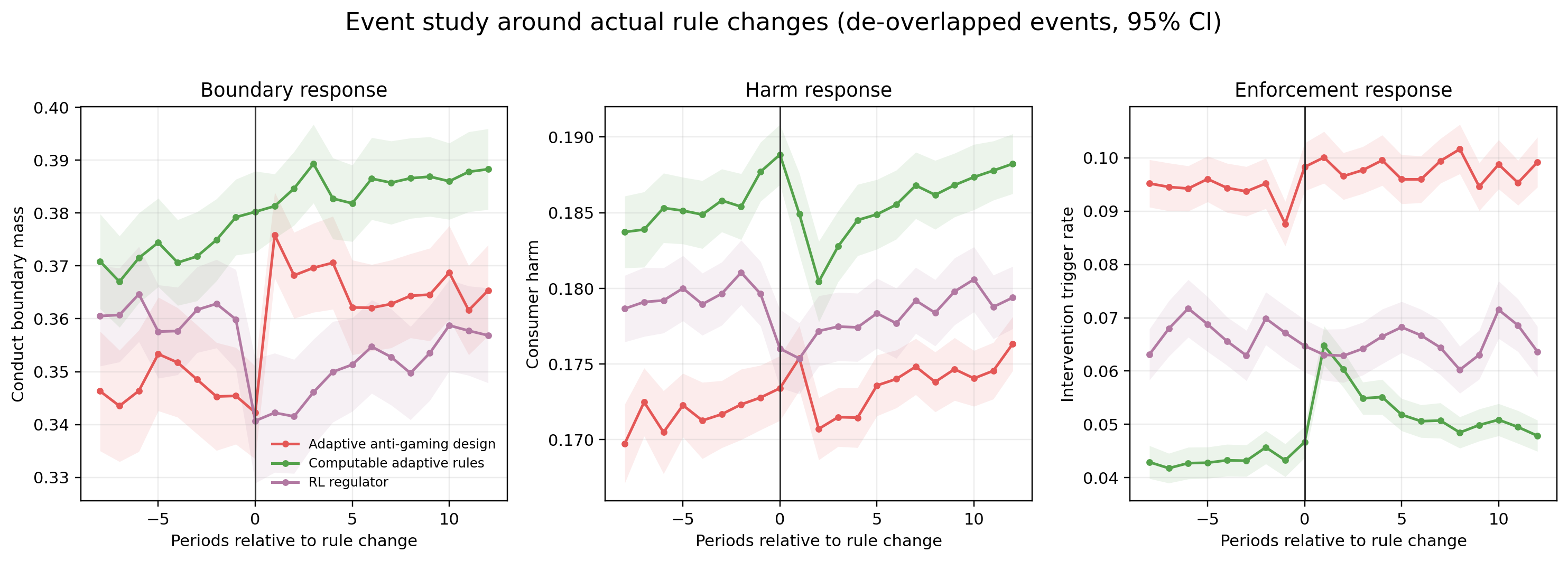}
\caption{Event study around actual rule changes after de-overlapping nearby updates.}
\label{fig:event}
\end{figure}

\begin{figure*}[t]
\centering
\includegraphics[width=\textwidth]{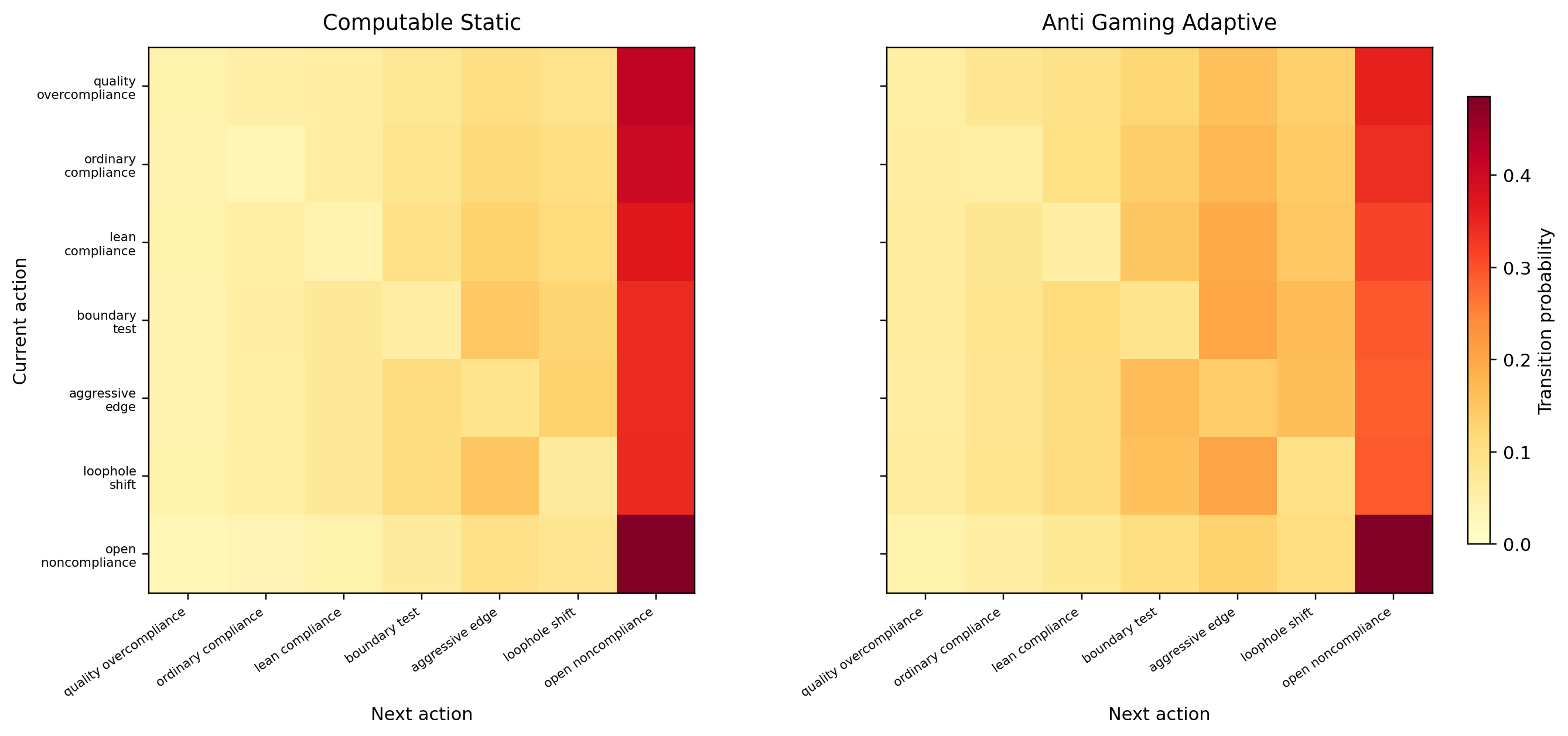}
\caption{Thirty-seed firm strategy transition matrices.}
\label{fig:transition}
\end{figure*}

\begin{figure}[t]
\centering
\includegraphics[width=\columnwidth]{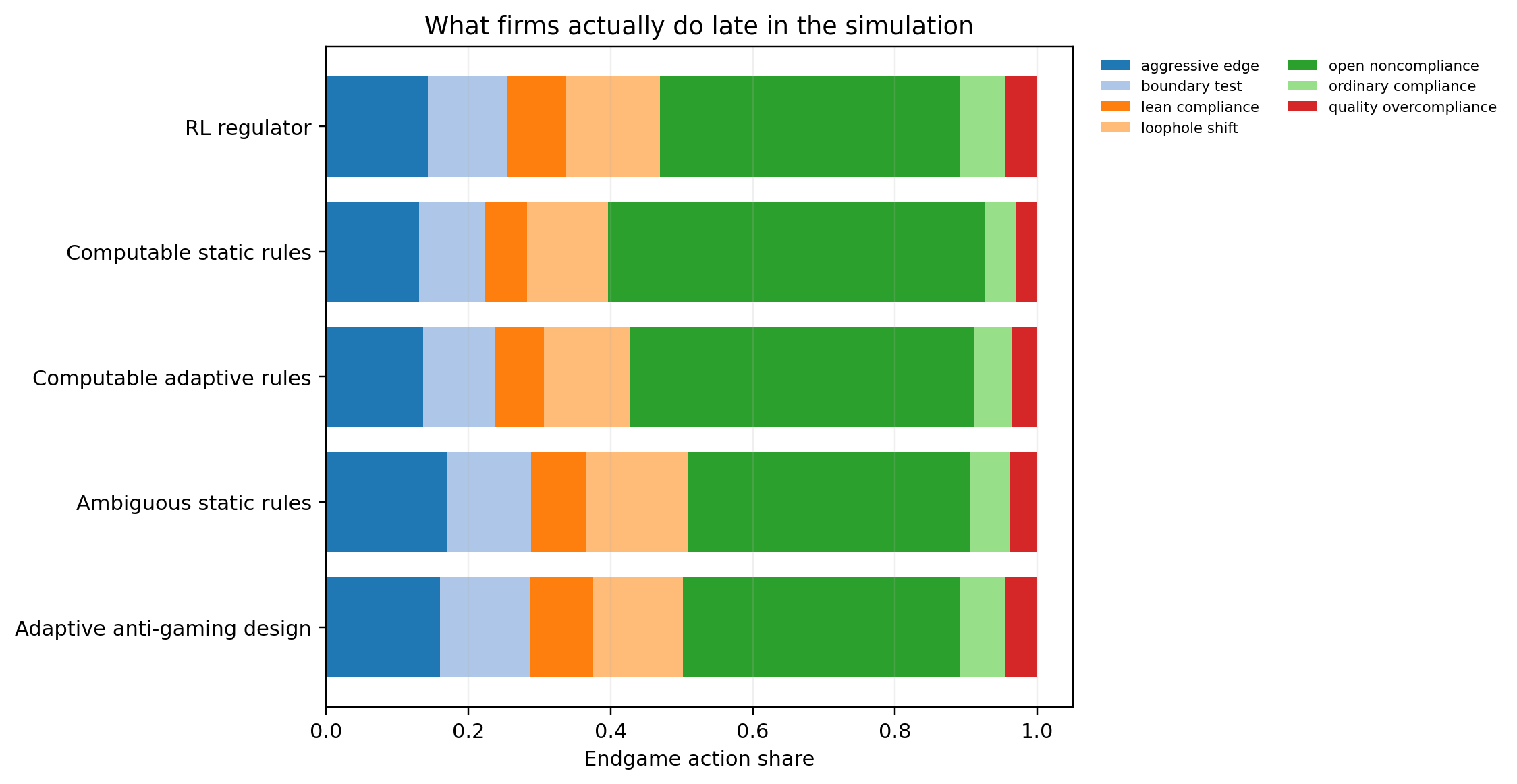}
\caption{Late-simulation firm action distribution across all panel seeds.}
\label{fig:endgame}
\end{figure}

\subsection{Statistical tests and diagnostic taxonomy}

The paired sign-flip tests are a check against overreading mean differences. Relative to computable static rules, anti-gaming adaptive design reduces conduct boundary mass by 0.032 ($p<0.001$) and consumer harm by 0.025 ($p<0.001$). These p-values quantify stochastic variation across simulation seeds under the specified model; they are not sampling uncertainty about real firms. The table also reports Holm-adjusted p-values because the same matched seeds are used for several outcomes. The taxonomy is deliberately downgraded: it is an internal diagnostic label generated from simulation metrics, not independent evidence that a real equilibrium type has been discovered.

\begin{table*}[t]
\centering
\caption{Paired tests relative to computable static rules. Intervals are paired bootstrap confidence intervals for the mean difference; p-values are two-sided sign-flip p-values. Holm-adjusted p-values correct across the paired-test family.}
\label{tab:tests}
\footnotesize
\setlength{\tabcolsep}{5pt}
\begin{tabular}{llrrrrr}
\toprule
Treatment & Metric & Mean diff. & 95\% low & 95\% high & $p$ & Holm $p$\\
\midrule
Computable adaptive & Boundary mass & -0.002 & -0.008 & 0.003 & 0.368 & 0.737\\
Computable adaptive & Consumer harm & -0.008 & -0.009 & -0.008 & $<0.001$ & 0.004\\
Computable adaptive & Edge strategy & -0.025 & -0.026 & -0.023 & $<0.001$ & 0.004\\
Computable adaptive & Formal violation & -0.048 & -0.052 & -0.043 & $<0.001$ & 0.004\\
Computable adaptive & Intervention trigger & 0.018 & 0.017 & 0.018 & $<0.001$ & 0.004\\
Anti-gaming adaptive & Boundary mass & -0.032 & -0.036 & -0.027 & $<0.001$ & 0.004\\
Anti-gaming adaptive & Consumer harm & -0.025 & -0.026 & -0.023 & $<0.001$ & 0.004\\
Anti-gaming adaptive & Edge strategy & -0.066 & -0.069 & -0.063 & $<0.001$ & 0.004\\
Anti-gaming adaptive & Formal violation & -0.060 & -0.065 & -0.054 & $<0.001$ & 0.004\\
Anti-gaming adaptive & Intervention trigger & 0.057 & 0.056 & 0.058 & $<0.001$ & 0.004\\
RL regulator & Boundary mass & -0.029 & -0.036 & -0.022 & $<0.001$ & 0.004\\
RL regulator & Consumer harm & -0.018 & -0.019 & -0.017 & $<0.001$ & 0.004\\
RL regulator & Edge strategy & -0.058 & -0.061 & -0.055 & $<0.001$ & 0.004\\
RL regulator & Formal violation & -0.086 & -0.093 & -0.079 & $<0.001$ & 0.004\\
RL regulator & Intervention trigger & 0.047 & 0.045 & 0.049 & $<0.001$ & 0.004\\
\bottomrule
\end{tabular}
\end{table*}

\begin{table}[t]
\centering
\caption{Diagnostic taxonomy across seed-level runs.}
\label{tab:taxonomy}
\footnotesize
\resizebox{\columnwidth}{!}{%
\begin{tabular}{@{}llr@{}}
\toprule
Scenario & Equilibrium type & Runs\\
\midrule
Ambiguous static rules & Boundary gaming equilibrium & 30\\
Adaptive anti-gaming design & Mixed adaptation & 30\\
Computable adaptive rules & Regulatory chase cycle & 30\\
Computable static rules & Boundary gaming equilibrium & 30\\
RL regulator & Regulatory chase cycle & 30\\
\bottomrule
\end{tabular}}
\end{table}

\section{Discussion}

The simulation does not show that computable regulation is undesirable. It shows why execution quality is an incomplete criterion. The same formalization that makes a rule easier to test also creates a more legible object for regulated firms to learn. That is a measurement problem as much as a compliance problem: regulators need to know whether a rise in apparent boundary clustering reflects a cleaner signal, a change in conduct, or both.

The policy implication is therefore design-oriented. Computable rules should report conduct-oriented diagnostics separately from signal-oriented diagnostics. Deterministic thresholds should be paired with randomized audit attention where the legal framework permits it, but randomization should select review or inspection attention rather than make hidden substantive law. Outcome guardrails should trigger review, safe-harbor withdrawal, burden shifting, or enhanced audit unless automatic liability is explicitly authorized. Each anti-gaming trigger should remain contestable through notice, reasons, recordkeeping, and review. These choices fit the responsive-regulation tradition, but they put more pressure on adversarial learning and on the epistemic limits of computable enforcement data \cite{ayres1992responsive,baldwin2008responsive}.

The analysis also qualifies the current enthusiasm for adapt-and-learn governance. Faster rule revision can become part of the game if firms learn the update logic. In the simulation, ordinary adaptation reduces harm but does not reliably reduce boundary mass. The stronger design is not simply the one that updates more often; it is the one that combines memory, budget accounting, randomized attention, and substantive outcome checks.

\section{Limitations}

The model is synthetic and deliberately stylized. It abstracts from any single jurisdiction, industry, or legal domain in order to isolate the boundary-learning mechanism. The action space is discrete, the Q-learning implementation is simple, and consumers observe risk only imperfectly. Legal institutions such as courts, procedural review, lobbying, political delay, heterogeneous legal advice, and strategic litigation are reserved for extensions because each could slow or redirect the feedback cycle studied here.

Computability enters through several channels at once: threshold clarity, enforcement-signal precision, adjustment speed, enforcement visibility, imitation, and loophole-harm scaling. I treat that bundle as institutionally plausible, while using the channel ablation to show how the modeled mechanism changes when individual channels are reset to the low-computability baseline. The Latin-hypercube design, common-random-number sweeps, boundary-epsilon checks, and penalty/reputation variants provide additional stress tests around the same mechanism.

The data package is synthetic as well. The firm-period panel provides transparent micro evidence about simulated agents, and the Q-learning regulator is intentionally simple so that it can serve as a diagnostic baseline rather than a tuned institutional design. Anti-gaming design reduces harm in this version, but it does not remove edge strategies; a richer model would distinguish formal loophole use, harmful loophole use, complaints, review outcomes, and legal challenges to adaptive triggers. The natural next steps are to calibrate the model to a concrete regulatory domain, add richer legal uncertainty, model judicial review of adaptive rules, and test whether observed firms cluster near real thresholds after digitization.

\section{Conclusion}

Making regulation computable changes the strategic environment, but the change is not captured by a single boundary-mass number. Once actual conduct is separated from enforcement signals, computable static rules can increase both real boundary clustering and observed signal clustering, with the signal response larger. Adaptive rules can reduce harm while adding rule churn, and budget-neutral anti-gaming design reduces harm and boundary mass without eliminating edge strategies. The paper's claim is cautious: computable regulation should be built not only for execution, but for adversarial learning, budget accounting, and careful measurement of what the new signals actually show.

\section*{Artifact Appendix}

The artifact is designed to make the mechanism inspectable and reproducible. It contains the simulation code, scenario parameters, synthetic data, seed-level outputs, generated figures, manuscript sources, and integrity metadata. A smoke-test reproduction runs with \texttt{.venv/bin/python run\_all.py --mode quick}. Full reproduction runs with \texttt{.venv/bin/python run\_all.py --mode full}; the release logs record 593.27 seconds in the project workspace and 580.19 seconds from a clean extracted package workspace using the same Python environment. The full firm-period panel is distributed as \texttt{data/firm\_period\_panel\_sample.csv.gz}, with a small CSV preview and panel metadata. The files \texttt{ARTIFACT\_MANIFEST.md} and \texttt{SHA256SUMS.txt} map source files, outputs, tests, and figures; the final package self-check passed, and the included test suite reports \texttt{10 passed}.

\end{document}